 \documentclass[preprint2]{aastex}

\slugcomment{Submitted to ApJ}

\shorttitle{$FWHM-K_2$ correlation in Black-Hole transients}
\shortauthors{Casares}

\begin{document}

\title{A $FWHM-K_2$ correlation in Black-Hole transients}

\author{J. Casares}
\affil{Instituto de Astrof\'\i{}sica de Canarias, 38205 La Laguna, 
S/C de Tenerife, Spain}
\affil{Departamento de Astrof\'isica, Universidad de La Laguna,
E-38206 La Laguna, S/C de Tenerife, Spain}

\begin{abstract}
We compare H$_{\alpha}$ emission profiles of 12 dynamically confirmed black holes 
(BHs) and 2 neutron star X-ray transients (SXTs) in quiescence with those of a sample 
of 43 Cataclysmic Variables (CVs), also quiescent. The full-width-half maximum 
($FWHM$) of the H$_{\alpha}$ line in SXTs is tightly correlated with the 
velocity semi-amplitude of the donor star $K_2=0.233(13)~FWHM$. 
This new correlation, when combined with orbital periods (i.e. through 
photometric light curves), opens the possibility to estimate  compact object 
mass functions from single integration, low-resolution spectroscopy. 
On the other hand, CVs above the period gap are found to follow a 
flatter correlation, a likely consequence of their larger mass ratios.   
We also find that the $FWHM$ traces the disc velocity at 
$\approx$42\% $R_{L1}$, independently of binary mass ratio. 
In addition, for a given $FWHM$, BHs tend to have lower $EWs$ than CVs. This might be 
explained by the fact that CVs must be seen at higher inclinations to mimic the 
same projected disc velocities as BH SXTs. 
For the same reason CVs with $FWHM\gtrsim1500$ km s$^{-1}$ 
are mostly eclipsing while none of our sample BHs are.
Further, we show that there is a vacant/unoccupied region for CVs in 
the $FWHM-EW$ plane defined by $FWHM > 2568~\sqrt{ ( 1 - (9/EW)^{2} )}$ (km/s). 
Both the $FWHM-K_2$ correlation and the $FWHM-EW$ plane can be exploited,
together with photometric light curves, to efficiently discover 
quiescent BHs in deep H$_{\alpha}$ surveys of the Galactic Plane. 
\end{abstract}


\section{Introduction}

Black holes (BH) provide unique laboratories to study key astrophysical 
phenomena such as accretion, the production of relativistic outflows and 
gamma ray bursts (e.g. \citealt{fender14, brown00}). BHs are usually 
discovered in transient X-ray binaries (SXTs) thanks to 
their dramatic, large-amplitude 
X-ray outbursts\footnote{By contrast, persistent X-ray binaries tend to harbour 
neutron stars.}. 
These are caused by thermal-viscous 
instabilities in accretion discs which are (slowly) fed by matter transferred 
from a (donor) companion star \citep{lasota01}.
The distribution of BH masses 
and binary periods provide fundamental constrains to models of 
supernovae explosions and compact binary evolution. Unfortunately, 
our understanding of the physics involved in these processes is 
still patchy because observational data are severely limited by small samples. 

The current sample of $\sim$50 BH SXTs, with only  
17 dynamical confirmations, is the tip of the iceberg of a hidden 
population of hibernating BHs (see \citealt{casares14a}).  
The size of 
the overall population is highly uncertain. Extrapolation of the number of SXTs 
discovered in the X-ray era suggests that several thousand dormant BHs are 
awaiting to be discovered \citep{vandenheuvel92,tanaka96,romani98}.
On the other hand, modern population-synthesis models 
predict an even larger population of $\sim10^4-10^5$ SXTs 
\citep{pfahl03,yungelson06}.
Observational constraints are likely biased low 
because they suffer from incompleteness and neglect a hitherto unexpected 
population of long period SXTs with very long duty cycles or even
supressed outburst activity (cf. \citealt{ritter02,menou99}). 
Furthermore, there is mounting evidence for the existence of a 
population of X-ray obscured or intrinsically faint BH SXTs  
\citep{corral13,armas14}.
It has been shown that the latter could be members of a sizeable population of
short period BH transients with low outburst luminosities \citep{maccarone13}.
And finally, the recent discovery of the first Be/BH binary indicates that 
some BHs might be accreting "silently" from the slowly outflow winds of 
rapidly rotating Be-type stars \citep{casares14b}. 
While binary population models predict a very modest number of Be/BH binaries in 
the Milky Way \citep{belczynski09,grudzinska15} empirical constraints, with only 
one detection yet, are very loose.  
Altogether, the only way to make significant progress in our knowledge of the 
Galactic population of BHs requires the discovery of a large sample of quiescent 
SXTs and this demands a new research methodology. 

Quiescent BH SXTs are particularly difficult to find since they are relatively 
faint across the electromagnetic spectrum in this state.  
By their very nature, quiescent states are 
characterised by extremely low accretion levels ($<10^{-9}$ M$_{\odot}$ yr$^{-1}$). 
The inner disc is truncated in an advected flow and, further, the lack of a 
solid surface (an exclusive signature of BH) results in extremely weak X-ray, 
UV and radio luminosities \citep{narayan08,miller-jones11}.
On the other hand, the optical spectrum 
is dominated by the low-mass donor star, with superposed emission lines from 
the accretion disc gas. The lack of a hard radiation field implies that only 
emission from neutral H and He is detected. Crucially, the strongest emission 
line is H$_{\alpha}$, but several other Galactic populations are also strong 
H$_{\alpha}$ emitters, such as Cataclysmic Variables (CVs), Symbiotic binaries, 
flare stars, reddened Be stars, T Tauri and other classes of young stellar
objects. As a consequence, H$_{\alpha}$ surveys of the Galactic plane are vastly 
outnumbered by these populations of contaminating H$_{\alpha}$ emitters. 
Attempts to clear out the sample using color selection cuts \citep{corral08} 
or X-ray diagnostics \citep{jonker11} still have to prove their effectiveness. 
Interestingly, new generation radio surveys (SKA and its pathfinders) 
may offer an alternative route to detect quiescent BHs given the increase in 
radio-to-X-ray flux ratios at very low luminosities \citep{maccarone05}. 

Here we present the discovery of a correlation between the width of the 
H$_{\alpha}$ line and the projected velocity of the companion star in quiescent 
SXTs. This property can be applied, in combination with photometric orbital
periods, to gather compact object mass functions and flag new potential BHs.
This strategy can be turned into a novel technique to unveil hibernating BHs, 
technique which appears much more efficient than traditional methods based on 
time-consuming spectroscopic monitoring of new X-ray novae.

\section{The sample}

\subsection{X-ray Transients} \label{SXTs}
We have assembled a spectroscopic database of dynamically confirmed BH SXTs 
with H$_{\alpha}$ emission (see Table 1). Most spectra were collected 
by us (V404 Cyg, BW Cir, N Mus 91, GS 2000+25, A0620-00, XTE J1650-500, XTE
J1859+226, GRO J0422+320, XTE J1118+480) and have been presented in several 
publications, 
while others were kindly provided by J. Orosz (XTE J1550-564), R. Remillard 
(N Oph 77) and A. Fillipenko (N. Vel 93). In addition, three new unpublished spectra 
of GRO J0422+320 were obtained on the night of 28 Jan 2009 with ALFOSC on the 
2.5m Nordic Optical Telescope (NOT) at the Observatorio del Roque de los Muchachos. 
Integration time was
set to 2400s and we used grism \#4 and 1" slit to cover the range
$\lambda\lambda$3820-9140 at 360 km s$^{-1}$ resolution. Seventy 
unpublished H$_{\alpha}$ spectra of V404 Cyg were also collected between 
1994-2005 with the 2.5m Isaac Newton Telescope (INT) and the 4.2m 
William Herschel Telescope (WHT) at resolutions 36$-$134 km s$^{-1}$. 
Five additional spectra of V404 Cyg were obtained on the nights of 5 and 7-8 
July 2008 at 200 km s$^{-1}$ resolution with the 2.1m telescope at the Observatorio San 
Pedro M\'artir (SPM). 
Finally the sample also includes spectra of the two only neutron star SXTs 
with a reported radial velocity curve of the companion star, 
Cen X-4 and XTE J2123-058. 
Table 1 provides the main observational details and associated references for
every system. 
   
\subsection{Cataclysmic Variables} \label{cv}

We also collected a database of H$_{\alpha}$ spectra of Cataclysmic Variables
(CVs) in quiescence including 41 dwarf novae and two intermediate polars, also 
classified as classical novae (GK Per and DQ Her). Most of the spectra were acquired 
during several observing campaigns and Service nights performed with the 
WHT in 1992, 1993, 1998 and 2006, the INT in 1992, 2001, 2008 and 2009, the NOT 
telescope in 2008, 2009 and 2012, the 2.2m telescope at Calar Alto Observatory in 1995 and the 2.1m 
telescope at SPM in 2008. The WHT spectra were obtained with ISIS and the R1200R grating which delivers 
40 km s$^{-1}$ resolution at H$_{\alpha}$. The INT 
spectra were obtained with the IDS spectrograph and gratings R150V, R300V, R900V and R1200R 
covering the H$_{\alpha}$ region at 50-320 km s$^{-1}$ resolution. 
For the NOT campaigns we used ALFOSC and grism \#7 with different slit widths 
providing 235$-$310 
km s$^{-1}$ resolution. The Calar Alto data were collected with the CASSPEC 
spectrograph, the f3 camera and grating \#11 in second order which delivered 
27 km s$^{-1}$ resolution. 
The SPM spectra were obtained with the Boller \& Chivens spectrograph and a 600 l/mm grating
 to yield 192 km s$^{-1}$ resolution. 
Spectra of 13 CVs (BV Cen, EY Cyg, DX And, HS0218+3229, AH Her, HS2325+8205, 
SDSS J100658.40+233724, U Gem, CTCV J1300-3052, OY Car, V2051 Oph,   
SDSS 103533.02+055158 and SDSS J143317.78+101123.3) were kindly provided 
by different authors  
while  
H$_{\alpha}$ parameters for other 15 CVs were compiled 
from literature. 
In addition, spectra of WZ Sge were retrived from the ING archive and reduced by us. 
The final collection of CV spectra were employed as a test sample for 
comparison with the H$_{\alpha}$ properties of SXTs. All the CVs have secured
velocity amplitudes ($K_2$) of their companion stars either through radial 
velocity studies or eclipse light curve solutions (for short period binaries 
where the companion star is undetected). 
Table 2 summarizes the main observational details of the CV database.
The database is intended to be a representative sample of quiescent CVs 
with available $K_2$ determinations although we warn about possible selection 
effects. These will be addressed in the following Section.

\section{The $FWHM-K_{2}$ correlation} 

Full-width-half-maximum ($FWHM$) values were obtained from gaussian fits to 
individual H$_{\alpha}$ profiles in every SXT. The fitted model consists of 
a constant plus a Gaussian function. Continuum rectified spectra were fitted 
in a window of $\pm$10000 km s$^{-1}$, centered on the H$_{\alpha}$ line after 
masking the neighboring HeI line at $\lambda$6678. We adopted 1-$\sigma$ formal 
errors on the fitted parameter as derived through $\chi^2$ minimization. Fig. 1 
displays some fit examples to average line profiles covering the entire range of 
line widths displayed by our data. It is clear from the figure that a simple Gaussian 
does not provide an accurate description of very broad profiles with large double peak 
separations. However, we find that the $FWHM$ given by the Gaussian model is within 
10\% of other width parameters obtained from more sophisticated double-Gaussian models. 
And, more importantly, it is far more robust since double-Gaussian models can 
easily fail when fitting low signal-to-noise profiles. In addition to the $FWHM$, we 
also extracted equivalent widths ($EW$) by integrating the H$_{\alpha}$ 
flux in individual spectra, after continuum normalization.  

Our 266 spectra of V404 Cyg 
span over 20 yr and, therefore, present the most complete database yet for the 
analysis of the 
secular 
evolution of accretion discs in quiescent BHs. Fig. 2 presents the evolution 
of both line parameters, $FWHM$ and $EW$,  
in V404 Cyg. The data points have been folded into 50 day  bins, except for the 
first 30 days where 1 day 
bins were used to better trace the rapid evolution through the outburst. 
The plot shows a steep rise in $FWHM$ (drop in $EW$) followed by a plateau 
phase starting $\sim$1300 days after the peak of the outburst. By taking the 
$FWHM$ as a proxy of the disc radius, 
Fig. 2 indicates that the accretion disc shrinks during  
outburst, 
recovering an equilibrium radius $\sim$3.5 yr after the maximum. 
Evidence of accretion disc shrinkage following outburst has been reported 
for CVs using eclipse timing and eclipse mapping techniques \citep{smak84,baptista01}. 
 Because we are interested here in comparing average properties of stable 
 quiescent discs in SXTs we decided to trim all data obtained within  
 $\sim$1300 days after the peak of the outburst. This leaves 127 useful spectra of V404 Cyg 
 (all since 1993), 94 of BW Cir and 3 of GRO J0422+320 (from 2009). 

 Table 3 lists the mean $FWHM$ and $EW$ values for SXTs, where the quoted 
 uncertainties represent 1 standard deviation in the distribution of individual 
 measurements. Therefore, our errors mostly reflect time variability in  
 line width and normalized flux.  Line variability is mainly caused by 
 aperiodic flares \citep{hynes02}  
or long-lived disc asymmetries (e.g. hot spots), 
 modulated with the orbital period. In the cases of N. Oph 77 and Vel 93 only 
 one phase averaged spectrum is available and, thus, we adopt 
 $\sigma_{FWHM}=0.1~FWHM$ to account for the average 10\% variability 
 displayed by the remaining binaries.  By the same token, we adopt 
 $\sigma_{EW}=0.22~EW$ for objects where only one spectrum is available, based 
 on the mean standard deviation in the $EW$s of the remaining  systems. 
  Instrumental resolution was subtracted quadratically 
  from every $FWHM$ value and, therefore, the H$_{\alpha}$ widths 
  quoted in Table 3 are intrinsic. Table 3 also provides fundamental 
binary parameters, chiefly the orbital period ($P$), the radial 
velocity semi-amplitude of the companion star ($K_2$) and the mass of the 
compact star ($M_1$) and inclination ($i$), when available, with their 
associated references. 
  
We also collected $FWHMs$ and $EWs$ from H$_{\alpha}$ lines in our  
sample of  CV spectra listed in Table 2. In the case of 14 eclipsing binaries 
(EM Cyg, EX Dra, HS 2325+8205, DQ Her, SDSS J100658.40+233724.4, 
U Gem, IP Peg, CTCV J1300-3052, 
HT Cas, OY Car, V2051 Oph, SDSS 103533.02+055158.3, 
WZ Sge and SDSS J143317.78+101123.3) 
only spectra obtained $\pm$0.05 phases 
away from the eclipse minimum were considered. As 
already mentioned, 
$FWHMs$ and $EWs$ values were obtained from the literature for 15 CVs. 
For these cases we adopt a 7\% error in $FWHM$ and 14\% error in $EW$, derived 
from the mean variability measured in the other 28 CVs.  
These values are listed in Table 4, together with determinations of 
the orbital period and $K_2$ velocities. 

In Figure 3 we compare $K_2$ versus $FWHM$ and it can be seen that these 
quantities are tightly correlated in SXTs, with a Pearson correlation 
coefficient $r=0.99$. A linear fit yields the following relation

\begin{equation}
K_{2}=0.233(13)~FWHM 
\end{equation}

\noindent
where both quantities are given in km s$^{-1}$. A second order polynomial fit 
is not justified since the constant coefficient is consistent with zero at 
1$\sigma$. In order to assess the error in $K_2$ estimated from the 
correlation we computed the difference with the true (dynamical) 
$K_2$ for our 14 SXTs. The values are found to follow a Gaussian distribution 
with standard deviation of 22 km s$^{-1}$. We therefore conclude that robust  
estimates of the $K_2$ velocity can be obtained from the width of the 
H$_{\alpha}$ line in SXTs. 
The uncertainty in the coefficient of the correlation was estimated 
through a Monte-Carlo simulation of 10$^4$ events, assuming that the difference 
between the model and true $K_2$ values follow a Gaussian distribution with 
$\sigma=22$ km s$^{-1}$.  

The $FWHM-K_2$ correlation is 
expected from basic equations. Assuming that $FWHM$  
is determined by gas velocity at a characteristic disc radius $R_{\rm W}$  

\begin{equation}
\left( \frac{FWHM}{2} \right) ^{2}= \frac{G M_{1}}{R_{\rm W}} \sin^2 i 
\end{equation}

\noindent
where $ M_{1}$ is the mass of the accreting star and $i$ the binary 
inclination. On the other hand, the companion's velocity is given by 

\begin{equation}
K_{2}^2=\frac{G M_{1}^2}{a \left( M_{1}+ M_{2} \right)} \sin^2 i 
\end{equation}

\noindent
with $M_2$ the mass of the companion star. Therefore, 

\begin{equation}
\left(\frac{K_{2}}{FWHM}\right)^2= \frac{R_{\rm W}}{4~a \left(1+q\right)}  
\end{equation}
 
\noindent
where $q=M_{2}/M_{1}$ is the mass ratio and $a$ the binary separation. If we now 
assume $R_{\rm W}= \alpha R_{\rm L1}$ (with $\alpha<1$) and use 
Eggleton's relation \citep{eggleton83} to remove $R_{L1}/a$ then

\begin{equation}
\frac{K_{2}}{FWHM}=\frac{\sqrt{\alpha f(q)}}{2}  
\end{equation}

\noindent
where 

\begin{equation}
f(q)= \frac{0.49 \left(1+q \right)^{-1}}{0.6 + q^{2/3} 
\ln \left(1+q^{-1/3}\right)}
\end{equation}
 
In the domain of SXTs, with $q\simeq0.05-0.15$, the 
dependence of Eq. 5 with $q$ is modest because 
$\sqrt{f(q)}$ varies between 0.77$-$0.69. 
For typical BH SXTs, with $q=0.1$,  
$K_{2}/FWHM= 0.36 \sqrt{\alpha}$ 
and hence, the empirical correlation, described by eq. 1, 
implies that the $FWHM$ of the H$_{\alpha}$ line in BH 
SXTs traces the disc velocity at about 41\% $R_{L1}$.  
And given that $\sim80\%$ of the line flux is contained within 
one $FWHM$ and that typical quiescent disc radii reach 
$\sim 0.5-0.6 R_{L1}$ \citep{marsh94, casares95b} we observe 
that the bulk of the H$_{\alpha}$ emission arises 
from the outermost regions of the accretion disc. 

Equation 1 stems from the fact that the Keplerian velocity field in the H$_{\alpha}$ disc 
provides a fundamental dimension scale of the binary. 
Similarly to the rotational broadening of the donor star $V \sin i$  
(see \citealt{wade88}), the {\it mean} accretion disc velocity (traced by the 
$FWHM$) 
scales with the donor star's velocity. Since  
both are projected velocities along the line of sight, the dependence on 
inclination cancels out. However, unlike $V \sin i$, 
this new correlation 
is only weakly dependent on binary mass ratio and hence results in a 
very tight linear regression. 
We also note that eq. 1 is more robust than a former 
empirical relation between $K_2$ and the double peak separation 
\citep{orosz94, orosz95} because the latter traces the tidal 
(outer disc) radius which can be strongly affected by disc asymmetries, the 
presence of S-wave distortions driven by hot-spots and anomalously low 
(sub-Keplerian) velocities  (e.g. \cite{north02}).    

At this point it is instructive to compare how the    
CVs distribute in the $FWHM-K_2$ plane. 
But prior to this it is important to asses the impact of possible selection 
effects. To start with, since the companion star is mostly detected above the 
period gap, our reference sample is, in principle, biased toward  
large mass ratios. The effect is further exacerbated by the fact 
that the CV population in the 3-4 h period range is dominated by SW Sex and 
Novalike stars, systems in permanent outburst  \citep{rodriguez-gil07}. This 
explains the paucity of CVs with periods $<0.17$ d in our list. To compensate 
for the deficit of binaries with small $q$ values we made an effort to 
incorporate short period CVs. These are, however, strongly 
skewed towards high inclinations because only eclipsing CVs can yield reliable 
$K_2$ values (through light curve modeling) when the companion star is not 
detected. Fig. 4 presents histograms of the distribution of orbital periods,
mass ratios and inclinations in our CV sample, clearly depicting these selection 
effects. Mass ratios and inclination values were compiled from the references listed 
in  Table 4 and \cite{ritter03}. 
In the figure we make a distinction between CVs above and below/within the period gap. 

Figure 4 shows that the distribution of mass ratios is clearly bimodal  and can be 
described by two Gaussians: 32 long-period CVs cluster at $q=0.63$ with $\sigma=0.2$ 
while 11 short-period CVs define a much narrow peak centered at $q=0.12$ and with 
$\sigma=0.07$. We note that this is overall similar to the distribution of mass ratios 
obtained from the entire CV data available in the Ritter's catalogue 
\citep{ritter03}. The figure also highlights the fact that our short period CVs 
are strongly biased toward high inclinations. By constrast, there seems to be no 
significant bias in the inclination of CVs above the gap. 
In view of this, we find justified to distinguish hereafter between short-period CVs
(i.e. below/within the gap) and long-period CVs (above the gap), with the latter not 
being strongly affected by selection effects.   

The bottom panel in Fig. 3 presents the location of the CVs in the $FWHM-K_2$ 
plane. The figure shows that long-period CVs display a similar regression to
that found for SXTs albeit flatter, i.e. for a given $K_2$,  H$_{\alpha}$ lines are 
systematically broader than in SXTs.  A linear fit yields 
$K_{2}=0.169(16)~FWHM$, 
relation which could be used to infer $K_2$ velocities for quiescent CVs above 
the period gap. We attribute the flatter slope of the CV correlation to their
comparatively larger mass ratios, which leads to smaller $R_{L1}$ and 
thus smaller disc radii (in binary separation units). 

If we now bring q=0.63 (i.e. the peak in the q distribution of long-period CVs) 
into eqs. 5-6 and set $K_2/FWHM=0.169$ from the empirical fit one obtains 
$\alpha=0.43$, 
in excellent agreement with what was found for SXTs. This implies that the 
$FWHM$ of the H$_{\alpha}$ line is always formed at about 42\% $R_{L1}$, 
irrespectively of the binary mass ratio. Despite the large spread in mass 
ratios the CV correlation appears quite narrow, a consequence of the very weak 
dependence of $K2/FWHM$ with $q$ for large $q$ values 
(see dotted lines in the bottom panel of Fig. 3). 

The group of short-period CVs, on the other hand, concentrate at large $FWHM$ values because of their 
high inclinations.  They are seen to depart from the trend defined by the long-period CVs, 
approaching the SXT correlation, a result of their small q-values. Although  the 
short-period CVs represent a very biased sample, they are of particular interest
because they define the upper limit in the $FWHM$ distribution of the CV population.

\section{The $EW-FWHM$ plane}

In a given system, the $FWHM$ of any emission line depends on the binary 
inclination, orbital period and the mass of the compact object. By 
bringing $K_2=(2 \pi a \sin i)/ P (1+q)$ into eq. 5 and using Kepler's 
Third Law we find

\begin{equation}
FWHM\propto~\frac{\sin i }{\sqrt{\alpha \left( 1+q \right)^{4/3} f(q)}} 
\left( \frac{M_1}{P}\right)^{1/3} 
\end{equation}

\noindent
Here, the dependence on mass ratio is extremely weak, with 
$\sqrt{(1+q)^{4/3} f(q)}$ varying in the range 0.80 to 0.69 for $q=0.05-1$. 
Therefore, we can safely assume 

\begin{equation}
FWHM \simeq A \left( \frac{M_1}{P}\right)^{1/3} \sin i
\end{equation}
  
\noindent
where $A$ is a constant that can be calibrated using $P$ and $FWHM$ values 
listed in Table 3, together with 
dynamical masses and inclinations available in the literature for 9 SXTs 
(also listed in Table 3). 
These yield a mean value $A=876\pm48$ 
km s$^{-1}$, when $M_1$ and $P$ are expressed in units of M$_{\odot}$ and days, 
respectively. 

On the other hand, the $EW$ of the H$_{\alpha}$ line 
depends on the binary geometry. For instance, 
\cite{warner86} showed that the $EW$ of H$_{\beta}$ in 
CVs increases with inclination because of the reduction of continuum brightness 
as the disc is seen at large inclinations. Therefore, 
to a first-order approximation 
one can assume 

\begin{equation}
EW \approx \frac{B}{\cos i}
\end{equation}

\noindent
where the constant $B=9\pm8$ \AA~has been calibrated using the distribution of 
$EW$s and inclinations listed in Table 3~\footnote{If we use instead reliable 
masses and inclinations reported in literature for a subset of 25 CVs from our  
Table 4 we find $A=897\pm46$ km s$^{-1}$ and 
$B=7\pm5$ \AA, in good agreement with the calibration obtained using SXTs.}. 

Fig. 5 displays our sample of SXTs and CVs in the $EW-FWHM$ plane. Here we have 
used open triangles to mark eclipsing CVs. 
BHs are a factor $\sim10$ more massive than white dwarfs or NS and
thus should possess, on average, wider H$_{\alpha}$ lines by a factor 
$\sim 2.2$.  
Instead, we observe that only one BH (XTE J1118+480) stands out clearly in the 
right side of the diagram with $FWHM>2500$ km s$^{-1}$. 
The remaining BHs are mixed up with CVs at lower $FWHM$ values because they 
either have long orbital periods (i.e. cases of V404 Cyg and BW Cir with 
$FWHM\sim1000$ km s$^{-1}$) or are viewed at lower 
inclinations. Fortunately, they cluster in the central part of the diagram 
between $FWHM\sim 1500-2500$ km s$^{-1}$, a region populated by eclipsing 
CVs. Since the latter are easily detected through 
deep ($\sim$2-3 mag) eclipses we conclude that non-eclipsing 
binaries with $FWHM\gtrsim1500$ km s$^{-1}$ are good candidates to host BHs.    
Incidentally, a population of (short-period) eclipsing BHs with very wide 
H$_{\alpha}$ lines would be expected in the right part of the diagram. 
Scaling from the eclipsing CVs we predict them to show a factor 
$\sim 2.2$ larger widths i.e. $FWHM\sim4200$ km s$^{-1}$.
As a matter of fact, the transient X-ray binary Swift J1357-0933 has been proposed 
as an extreme inclination BH, although no dynamical solution is yet available 
\citep{corral13}. 
We have recently obtained 4 H$_{\alpha}$ spectra of Swift J1357-0933 with 
OSIRIS and the R300R grism on the 10.4m Gran Telescopio Canarias (GTC) 
on 29-30 June 2013, from which we measure $FWHM=4085\pm328$ km s$^{-1}$ and 
$EW=131.9\pm14.5$. These values are indeed consistent with a short period 
BH seen at very high inclination. The position of Swift J1357.2-0933 is 
indicated by an asterisk in the diagram.    
 
Using eqs. 8-9 one can define regions of constant $M_1/P$ and inclination in the
$EW-FWHM$ plane. These are marked in the bottom panel of Fig. 5 using dashed
and dotted lines respectively.  
We stress, however, that the quoted inclinations must be considered as mere 
indicative given the 
crude approximation involved in eq. 9. The figure 
suggests (with all the caveats associated to low number statistics) 
that, for a given $FWHM$, BHs tend to have lower $EWs$ than CVs. 
This could be explained because, due to their  
shallower potential wells, CVs must be seen at higher inclinations to mimic the 
same projected disc velocities as BHs and, therefore, their H$_{\alpha}$ fluxes 
are less diluted by the accretion disc continuum. 
It is interesting to note the position of GRO J0422+320 in the upper left side of 
the diagram. While the mass of its BH is quite uncertain  
(see \citealt{casares14a}) both the large $EW$ and low $M_{1}/P$ factor 
suggest it hosts a low-mass BH, in line with the results of 
\cite{casares95a} and \cite{gelino03}.   

Finally, we can tentatively define a forbidden region for CVs in the $FWHM-EW$ 
plane by taking extreme parameters i.e. $M_{1}>1.4$ M$_{\sun}$ and $P<80$ min, 
the period minimum spike observed in the distribution of CV periods 
\citep{gansicke09}. This yields $M_{1}/P>25.2$ M$_{\odot}$/d and thus   
$FWHM> 2568 \sqrt{ ( 1 - (9/EW)^{2} )}$, limit which is marked by a 
green solid line in the plot. 
As a test we have examined a random sample of 236 dwarf novae selected from 
Sloan DR7. Sloan CVs show the same trend as seen in Fig. 5 i.e. they populate 
the region leftwards of the green line, with a large spread towards high 
$EW$s up to $\sim$350 \AA .

\section{Discussion: new strategies to detect dormant BHs}

To make progress in our understanding of the formation and evolution of 
Galactic BHs it is essential to discover a large 
sample of secured (dynamically confirmed) BHs. This new sample would allow 
us to constrain the number density, the orbital period distribution and, 
ultimately, the BH mass spectrum. Only then we will be able to address fundamental 
questions such as the role of supernova models in shaping the distribution of BH masses, 
a current hot topic in the community \citep{ozel10, farr11, kreidberg12, belczynski12}.
Deep H$_{\alpha}$ surveys of the Galactic plane, combined with 
spectroscopic surveys,  
can efficiently select samples of H$_{\alpha}$ emitting objects with $FWHM>1000$ km s$^{-1}$.   
This width cut would allow instant removal of Galactic populations of 
narrow H$_{\alpha}$ emitters such as planetary nebulae, 
Be, chromospheric stars, T Tauri and other YSOs. Only high inclination CVs, due to the large gravitational 
fields of their accreting white dwarfs, are able to produce wide H$_{\alpha}$ 
lines. 

Fortunately, as 
we have shown, the width of the H$_{\alpha}$ line in quiescent BH and 
NS SXTs is tightly correlated 
with the projected velocity of the donor star. The $FWHM-K_{2}$ 
relation can, therefore,  be exploited, together with supplementary information 
on orbital periods (e.g. from light curve variability), to gather "preliminary" 
mass functions ($PMF$) of compact objects from single epoch spectroscopy 
i.e. $PMF=1.3\times10^{-9}~P~(FWHM)^3$, where $P$ is given in days and 
$FWHM$ in km s$^{-1}$.  
$K_2$ values can be estimated from single integrations rather than expensive 
time resolved spectroscopy, allowing a search for dynamical BHs in much  
deeper fields and using a factor $\sim$4 lower spectral resolution 
than usually employed. 
Therefore, the novel strategy that we propose is clearly much more efficient than 
standard spectroscopic techniques, aiming at measuring the orbit of the donor star 
from the Doppler shift of weak absorption lines. 
And it may be the only way to infer mass functions in extremely faint BH 
SXTs i.e. the bulk of the Galactic population. 
We also note that our technique can be easily executed in crowded regions like 
globular clusters, where HST time-series photometry can yield orbital periods 
while ground-based spectroscopy is severely limited by seeing conditions. 

Fig. 6 demonstrates how BHs are nicely segregated from CVs and NS 
using the $FWHM-K_2$ correlation. 
Every BH (except GRO J0422+320) is located in the right hand part of the diagram, beyond the dotted 
vertical line. 
Therefore, targets with $PMF>2$ M$_{\odot}$
are strong candidates to host 
BHs. 
It should be noted that the plotted CV $PMFs$ are, in most cases, 
robust upper limits to true mass functions because, by adopting eq. 1 we 
overestimate their real $K_2$ values. 
By comparing the $PMF$ values with the true (dynamical) mass function of BHs 
one can estimate the typical uncertainty introduced by the $FWHM-K_2$ 
correlation. 
Relative errors in mass function are found to follow an 
approximate Gaussian distribution with $\sigma=0.08$.  
Therefore, the $FWHM-K_2$ correlation allows us to estimate 
mass functions with typically  $\sim$10\% uncertainty. 
As an example, we have applied eq. 1 to 
Swift J1357.2-0933, where our quiescent GTC spectra yield 
$FWHM=4085\pm328$ km s$^{-1}$. 
Extrapolating from eq. 1 we predict $K_2=952\pm93$ km s$^{-1}$ which, when 
combined with $P=0.117\pm0.013$ d, \citep{corral13} leads to a record mass 
function $PMF=10.5\pm 3.3$ M$_{\odot}$. We note that this figure might even 
be slightly underestimated since the GTC spectra were taken only $\sim900$ d. 
after the peak of the outburst (see Sect. 3). In any case, the effect would be
small compared to our errorbar, which is dominated by propagating the large
$FWHM$ uncertainty (driven by line variability) into the $FWHM-K_2$ relation 
and the mass function equation.

A big step forward in the exploitation of the $FWHM-K_2$ 
relation will likely come from the use of imaging techniques. Accurate H$_{\alpha}$ 
widths can be measured directly through a combination of customized narrow-band 
filters, removing the need for any spectroscopy at all. This brings in 
a new observational signature that we coin here {\it Photometric Mass 
Function} ($PMF$, taking advantage from the same acronym before). $PMF$s open 
a novel concept, i.e. that of weighting mass functions photometrically, and 
lay the ground for the efficient discovery of new hibernating BHs in large 
survey volumes. We are currently working on this strategy. 

Finally, it should be borne in mind that there is a strong selection effect 
against detecting high inclination BHs in X-ray selected samples. This is 
because flared accretion discs 
obscure X-rays and indeed none of the currently known dynamical BHs has an  
inclination  $>75^{\circ}$ \citep{narayan05}. 
Since a $PMF$-based sample will be selected by H$_{\alpha}$ widths and not X-ray 
emission we expect this strategy will uncover a significant number of high inclination and 
even eclipsing BHs. These hold the prospect to render the most accurate BH masses yet because 
the relative uncertainty in the inclination dominates the mass error budget. 
Clearly, the newly discovered BHs will have an strategic impact in the 
construction of the BH mass spectrum. 

\acknowledgments
We would like to thank the anonymous referee for the valuable comments which 
helped to improve the quality of the paper. 
We also acknowledge the hospitality of the Department of Astrophysics of the Oxford 
University (UK) where part of this work was carried out. We are grateful to R. 
Remillard, A. Filippenko and J. Orosz for providing us
with spectra of N. Oph 77, N. Vel 93 and XTE J1550-564 respectively. Also to 
C. Watson, A. Bruch, K. Horne, C. Savoury, D. Steeghs, C. Copperwheat and 
S. Tulloch for the spectra of BV Cen, DX And, AH Her, CTCV J1300-3052, V2051 Oph, 
OY Car and SDSS J143317.78+101123.3, respectively. 
P. Rodr\'\i{}guez-Gil kindly provided spectra of HS0218+3229 and HS2325+8205 while 
J. Southworth those of SDSS J100658.40+233724 and SDSS 103533.02+055158.3. 
A selection of U Gem and EY Cyg spectra, the latest 
corrected for nebular emission, was provided by J. Echevarr\'\i{}a. 
We thank R. Cornelisse for taking the GTC spectra of Swift J1357-0933. 
We are indebted to C. Zurita for conducting the San Pedro M\'artir observations 
and also to the team of IAC Support Astronomers for 
obtaining CV spectra during Service time over several years. 
This paper makes use of data obtained from the Isaac Newton Group Archive 
which is maintained as part of the CASU Astronomical Data Centre at the 
Institute of Astronomy, Cambridge.
This work is supported by DGI of the Spanish Ministerio de Educaci\'on, Cultura y 
Deporte under grants AYA2010-18080, AYA2013-42627 and SEV-2011-0187. 
Partially based on observations made with the GTC operated on the island of La Palma 
by the Instituto de Astrof'sica de Canarias in the Spanish Observatorio del Roque de Los Muchachos
of the Instituto de Astrof'sica de Canarias.

\clearpage

\begin{figure}
\includegraphics[angle=-90,scale=0.55]{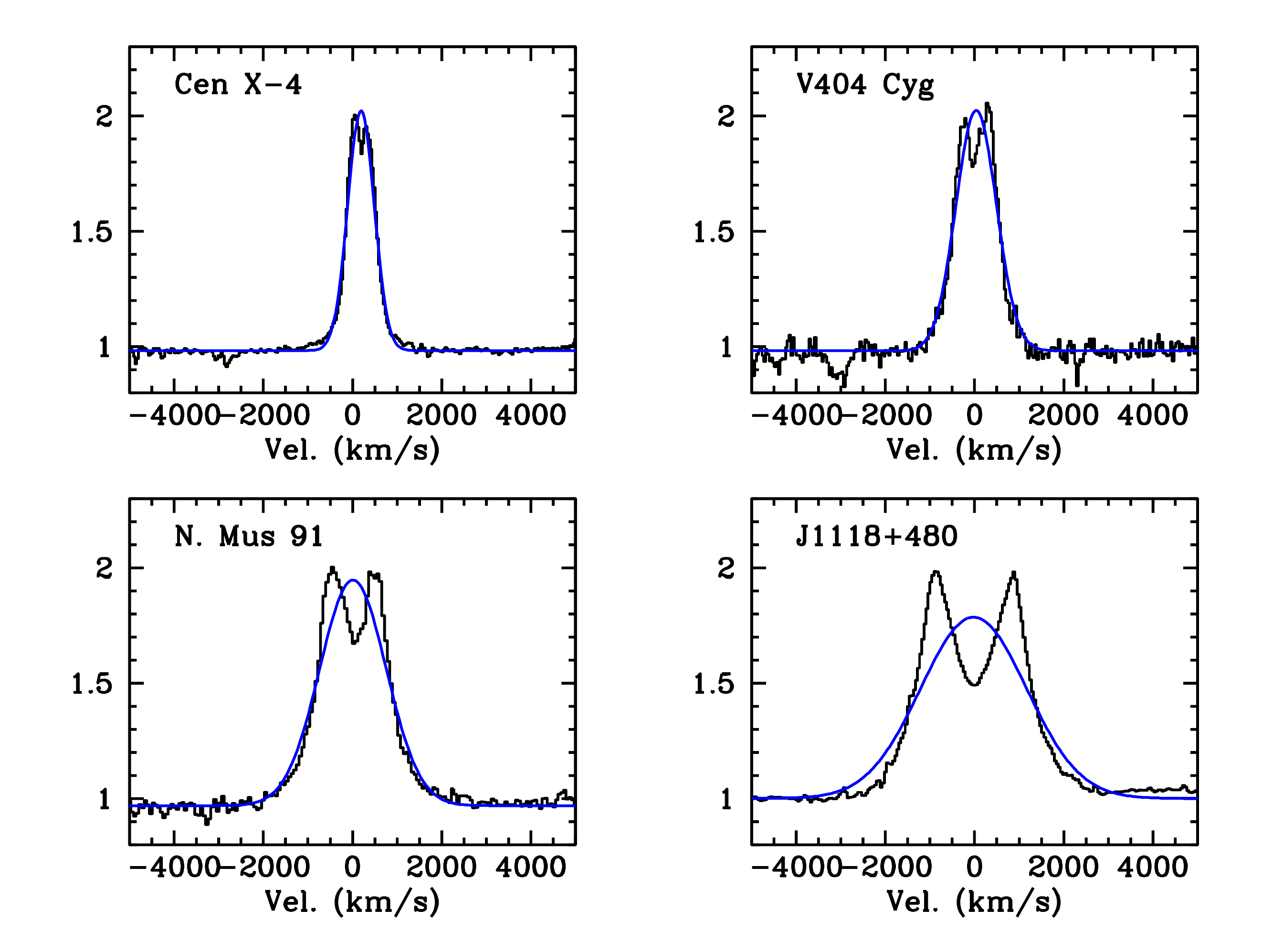}
\caption{Example of Gaussian fits to H$\alpha$ profiles in SXTs. A selection of average spectra, representing the entire range of  
$FWHM$s, is depicted. 
\label{fig1}}
\end{figure}

\clearpage

\begin{figure}
\includegraphics[angle=-90,scale=.50]{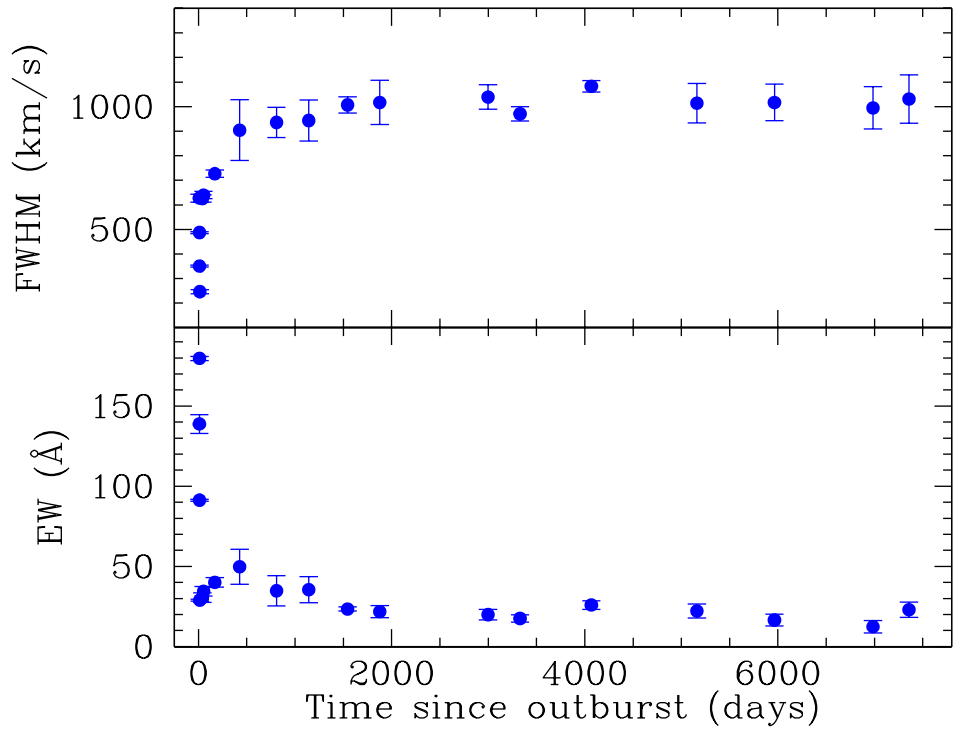}
\caption{Long-term (20 year) evolution of the $FWHM$ and $EW$ of the H$_{\alpha}$ 
line in V404 Cyg. Individual points have been co-added into 50 day bins
except for the first 9 data points where 1-day bins were used. 
Errorbars reflect variability within the bin. 
\label{fig2}}
\end{figure}

\clearpage

\begin{figure}
\includegraphics[angle=0,scale=0.55]{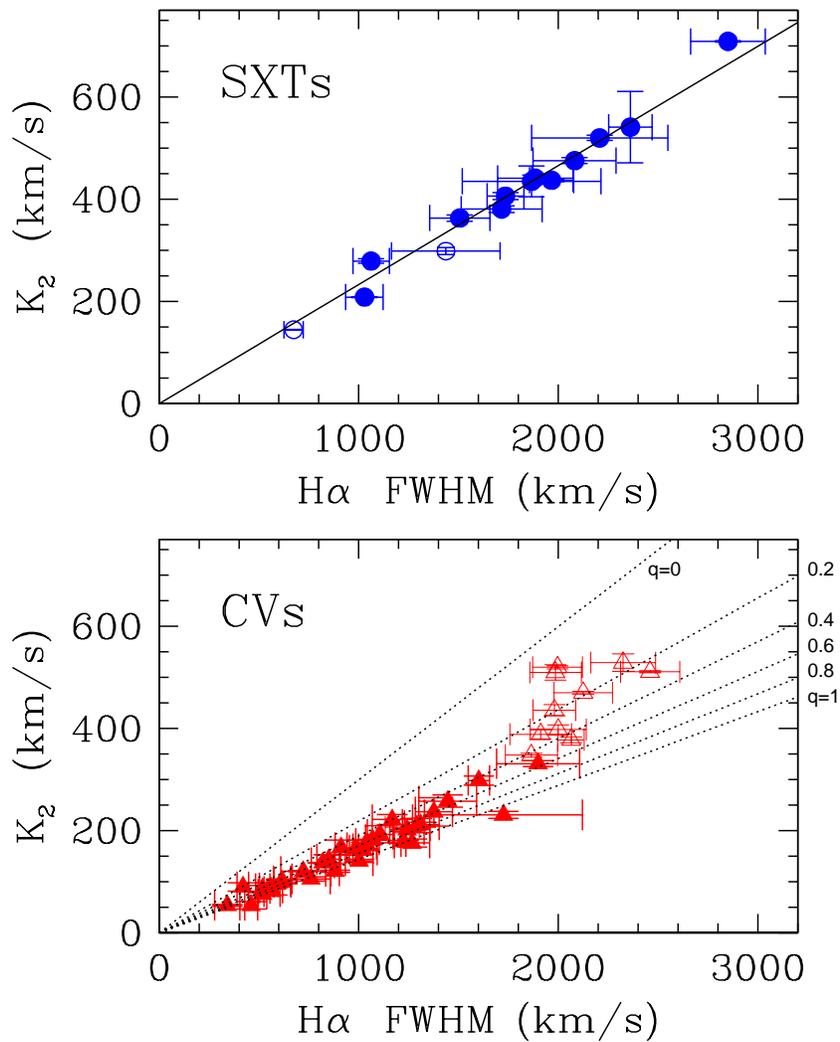}
\caption{Top: the $FWHM-K_2$ correlation for SXTs with the best linear fit. 
Blue solid circles indicate BHs while Ns are marked by open circles 
Bottom: the $FWHM-K_2$ correlation for  CVs. Filled triangles indicate CVs above the period gap while 
open triangles those below/within the gap. Like in the case of SXTs, long period CVs display a tight 
$FWHM-K_2$ correlation although with a flatter slope. Dotted lines mark theoretical correlations from 
eq. 5-6 for $q=0-1$ and $\alpha=0.43$. 
\label{fig3}}
\end{figure}

\clearpage

\begin{figure}
\includegraphics[angle=0,scale=.60]{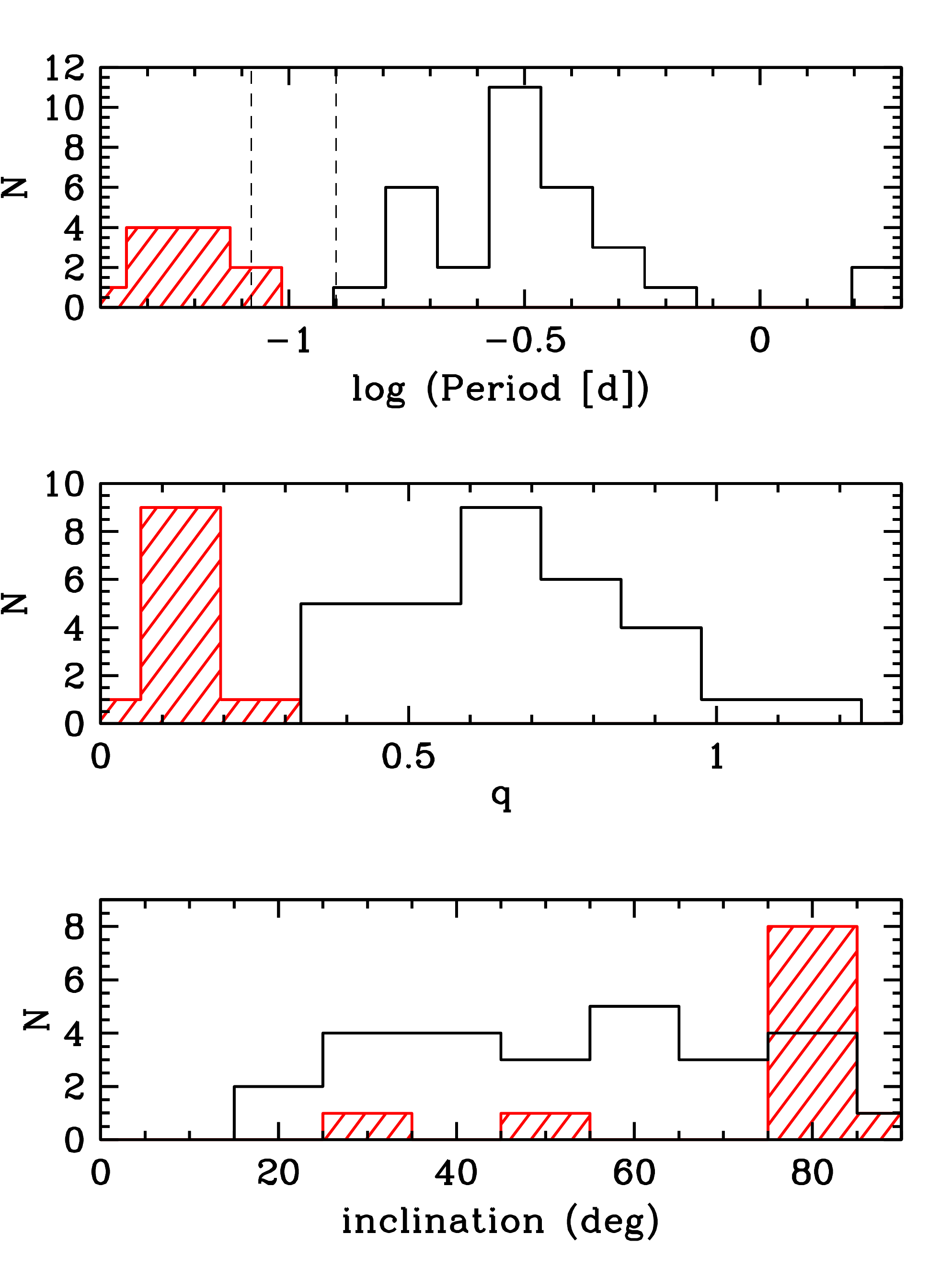}
\caption{Distribution of orbital periods (top panel), mass ratios (middle) and inclinations (bottom) 
for the test sample of CVs. The white histogram represents CVs with orbital periods 
above the period gap while the red shaded histogram those below/within the gap. The period 
gap is indicated by dashed vertical lines in the top panel. All the CVs from the white histogram  
possess $K_2$ determinations based on radial velocity curves of the donor stars  while 
those in the red shaded histogram on modeling eclipsing light curves. The only exceptions to the 
latter are the ultra-short period system EI Psc  and the period gap CV  
SDSS J1300-3052.  It can be seen that short period CVs have very low mass ratios and are 
strongly biased toward high inclinations (with the exception of the 2 CVs just mentioned) . 
\label{fig4}}
\end{figure}

\clearpage

\begin{figure}
\epsscale{.80}
\includegraphics[angle=0,scale=.60]{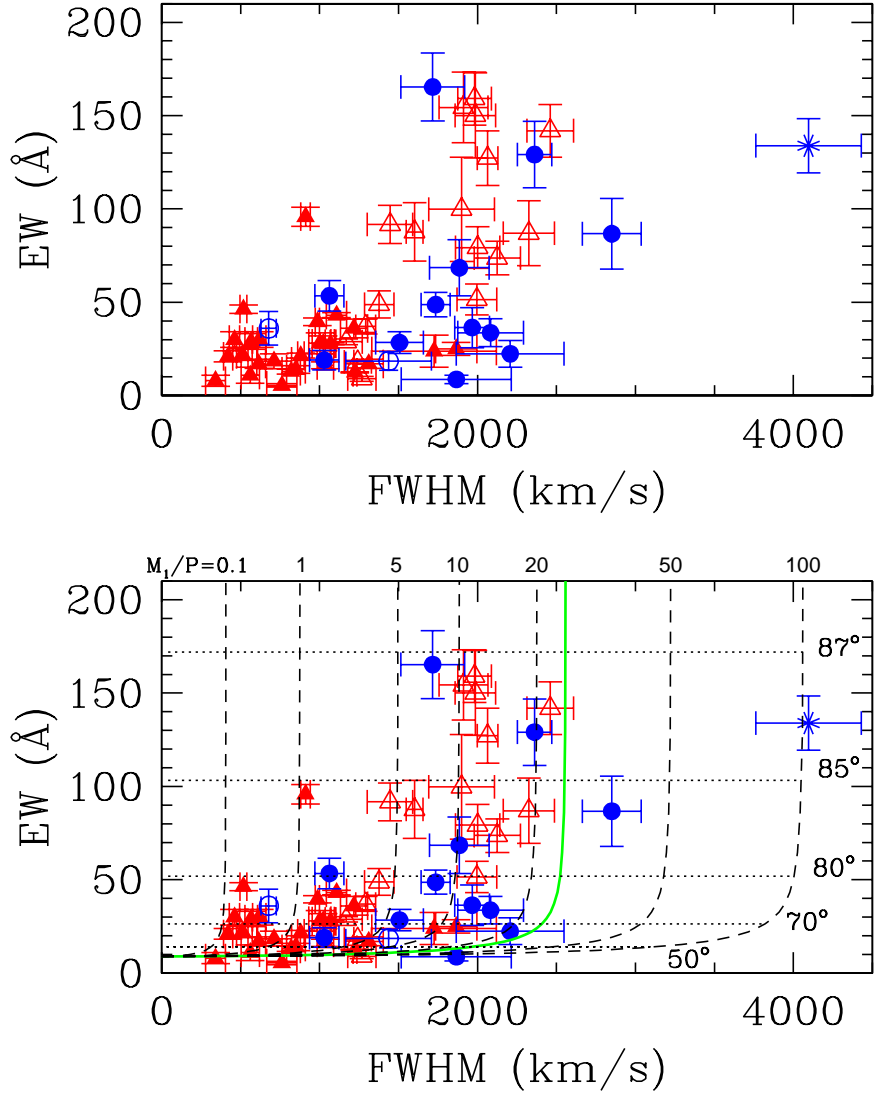}
\caption{Top: SXTs and CVs in the $FWHM-EW$ plane. Same symbol code is used as in 
Fig. 3, except for open triangles which now indicate eclipsing CVs. 
Bottom panel: dashed vertical lines correspond to constant $M_{1}/P_{\rm orb}$ 
values (expressed in units of M$_{\odot}$/d) while horizontal dotted lines 
indicate various inclinations, in an idealized case where $EWs$ are 
fortshortended by a factor $\cos i$ due to disc continuum visibility. The 
green solid line defines  the maximum $FWHM$ predicted for CVs.
\label{fig5}}
\end{figure}

\clearpage

\begin{figure}
\includegraphics[angle=-90,scale=.50]{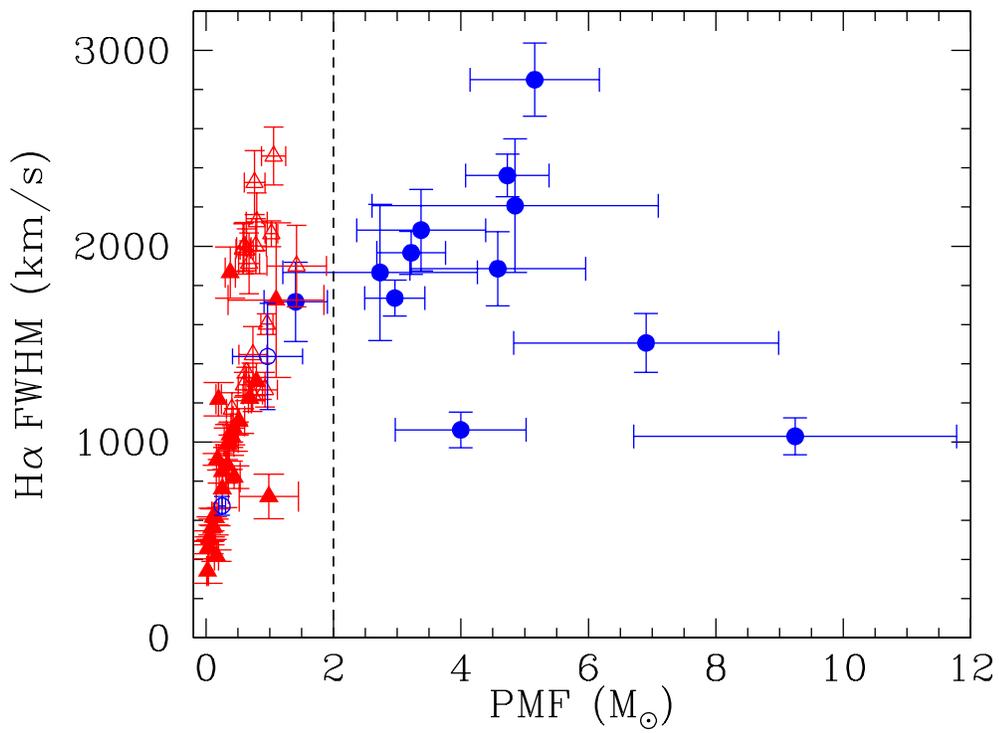}
\caption{Preliminary mass functions obtained from $K_2$ values derived through 
the $FWHM-K_2$ correlation for SXTs.  Same symbol code is used as in Fig. 5. 
\label{fig6}}
\end{figure}

\clearpage

\begin{deluxetable}{lcccc}
\tablewidth{0pt}
\tablecaption{Database of X-ray transients\label{tbl-1}}
\tablehead{
\colhead{Object} & \colhead{\# Spectra} & \colhead{Year} & \colhead{Resolution (km/s)} & \colhead{ref} } 
\startdata
\cutinhead{Black Holes}
V404 Cyg             & 266 &   1989-2009  & 6-180     & 1-8 \\
BW Cir                 &   96  &   1995-2006   &   70-110    & 9-10 \\
XTE J1550 -564  &     33  &    2001, 2008         & 55-165   & 11-12 \\
N. Oph 77            &     1   &   1993   &  180  & 13 \\
N, Mus 91            &   29   & 1993-1995    &   74-90  & 14 \\
GS 2000+25        &   25   &   1995           &  196 & 15 \\ 
A0620-00             &   20  &    2000          &      7 & 16 \\ 
XTE J1650-500    &  15   &  2002       &   35 & 17 \\ 
N Vel 93               &    1   &  1998   &   120   & 18 \\
XTE J1859+226   &  10   &  2010  & 255 & 19 \\
GRO J0422+320  &   21  & 1994-1995, 2009  & 230-630 & 8, 20 \\
XTE J1118+480    &  120  &  2011  & 120 & 21 \\ 
\cutinhead{Neutron Stars}
Cen X-4               &  90 & 1993-2002    & 6-74 & 22-24 \\ 
XTE J2123-058   &  20 & 2000 & 123-183 & 25 \\ 
\enddata
\tablerefs{
(1) \cite{casares91}; (2) \cite{casares92a}; (3) \cite{casares92b}; 
(4) \cite{casares93}; (5) \cite{casares94}; (6) \cite{hynes02}; (7) 
\cite{gonzalez11}; (8) This paper; (9) \cite{casares04}; (10) \cite{casares09a}; 
(11) \cite{orosz11};  (12) \cite{orosz02}; (13) \cite{remillard96}; (14) \cite{casares97}; (15) 
\cite{casares95b}; (16) \cite{gonzalez10}; (17) \cite{orosz04}; (18) 
\cite{filippenko99}; (19) \cite{corral11}; (20) \cite{casares95a}; (21) 
\cite{gonzalez12};(22) \cite{torres02}; (23) \cite{davanzo05}; (24) 
\cite{casares07}; (25) \cite{casares02}
.}
\end{deluxetable}

\clearpage

\begin{deluxetable}{lcccc}
\tablewidth{0pt}
\tablecaption{Database of Cataclysmic Variables.\label{tbl-2}}
\tablehead{
\colhead{Object} & \colhead{\# Spectra} &\colhead{Year} & \colhead{Resolution (km/s)} 
& \colhead{ref} }
\startdata
GK Per                &  8 &  1992, 1995, 1998  & 25-70 &  1-2 \\ 
SDSS 2044-04          & -  & 2001-2004  & 190-210 & 3 \\
BV Cen                & 63 & 2004 & 10 & 4  \\   
RX 1951.7+3716        & -  & 2001-2004  & 190-210 & 3 \\
UY Pup                & -  & 2002-2003   & 190-210 & 5 \\
EY Cyg                & 103  & 1998-2001, 2004  & 15-20 & 6 \\
DX And                &  4 & 1993  & 180  & 7 \\
SY Cnc                &  28 & 1992, 1998-1999, 2003, 2008  & 25-70 &1,8 \\
AT Ara                & -  & 2000 & 100 & 9 \\
RU Peg                &  2 & 2008 & 60 & 2 \\
GY Hya                & -  & 2001-2004  & 190-210 & 3 \\
CH UMa                & 16  & 2008-2009  & 60-320 & 2 \\
V392 Hya              & -  & 2001-2004  & 190-210 & 3 \\
RY Ser                & -  & 2003 & 190-210 & 5 \\
HS 0218+3229         & 56  & 2002  & 55-75 & 10 \\
EM Cyg                & 3 & 2008 & 60 & 2 \\
Z Cam                 &  12 & 2009 & 60 & 2 \\
SDSS 0813+45          & - & 2002 &190-210 & 5 \\
V426 Oph              & 4 & 2006 & 40 & 2 \\
SS Cyg                & 23 & 2008 & 60-250 & 2 \\
LY UMa                 & - & 2000 & 190-500 & 11 \\
BF Eri                  & 5 & 2009 & 235 & 2 \\
TT Crt                & - & 2000-2003 & 190-210 & 5 \\
AH Her                & 10 & 1980-1981 & 60  & 12 \\
EX Dra                & 31 & 2001 & 50 & 2 \\
HS 2325+8205          & 42 & 2005, 2007 & 215 & 13 \\
DQ Her                & 4 & 2009 & 60 & 2 \\
SDSS J100658.40+233724 & 68 & 2008 & 140 & 14 \\
SS Aur                &  28 & 2008, 2013 & 90-320  & 2 \\
U Gem                 & 32 & 1999, 2008 & 16-320 & 2,15 \\
CN Ori                & 17 & 1986 & 137 & 16 \\
IP Peg                & 28 & 1988, 2009 & 50-150 & 2,17 \\
CTCV J1300-3052       &  24 & 2010 & 45 & 18 \\
QZ Ser                & - &  2002  & 200 & 19 \\
Z Cha                 & - & 1984 & 150 & 20  \\
HT Cas                & 10 & 2008  & 60-190 & 2 \\
OY Car                &  28 & 2010 & 46 & 21 \\
V2051 Oph             & 31 & 1998, 2009 & 55-430 & 2,23 \\
SDSS 103533.02+055158.3 & 51 & 2006 & 70 & 23 \\
WZ Sge                & 384 & 1996 & 25 & 24 \\
SDSS J143317.78+101123.3 & 38 & 2008 & 32 & 25 \\
SDSS 1507+52          & - & 2006 & 160 & 26 \\
EI Psc = J2329+0628   & - & 2001 & 180 & 27 \\
\enddata
\tablecomments{Information on SDSS 2044-04, RX 1951.7+3716, UY Pup, AT Ara, 
GY Hya, V392 Hya, RY Ser, SDSS 0813+45, LY UMa, 
TT Crt, CN Ori, QZ Ser, Z Cha, SDSS 1507+52 
and EI Psc has been extracted from literature. 
In the case of CN Ori, the $FWHM$ and $EW$ values are obtained from 17 individual measurements listed in table 1 
of \cite{barrera89}. 
The AH Her spectra are phased binned averages. 
}
\tablerefs{
(1) \cite{martin95}; (2) this paper; 
(3) \cite{peters05}; 
(4)  \cite{watson07}; (5) \cite{thorstensen04}; 
(6)  \cite{echevarria07}; (7) \cite{bruch97}; (8) \cite{casares09b}; (9) \cite{bruch03}; 
(10) \cite{rodriguez-gil09}; (11) \cite{tappert01}; 
(12) \cite{horne86}; (13) \cite{pyrzas12}; (14) \cite{southworth09}; 
(15) \cite{echevarria07};  
(16) \cite{barrera89}; (17) \cite{harlaftis94}; (18) \cite{savoury12}; (19) \cite{thorstensen02a};  
(20) \cite{marsh87}; (21) \cite{copperwheat12}; (22) \cite{steeghs01}; (23) \cite{southworth06}; 
(24) \cite{skidmore00}, (25) \cite{tulloch09}; (26) \cite{patterson08}; (27) \cite{thorstensen02b}
 .}
\end{deluxetable}

\clearpage

\begin{deluxetable}{llrrrrrc}
\tablewidth{0pt}
\tablecaption{X-ray transients\label{tbl-3}}
\tablehead{
\colhead{Object} & \colhead{$P$ (d)}  & \colhead{$K_{2}$ (km/s)}  &
\colhead{$FWHM$ (km/s)} & \colhead{$EW$ (\AA)} & \colhead{$M_1$ (M$_{\odot}$)} & \colhead{$i$ (deg)} & \colhead{ref~\tablenotemark{\dagger}} }
\startdata
\cutinhead{Black Holes}
V404 Cyg        & 6.47129   &  208.4 $\pm$ 0.6 & 1029 $\pm$ 94  &19.0 $\pm$ 5.2  & 
$9.0^{+0.2}_{-0.6}$ &  $67^{+3}_{-1}$ & 1,2 \\
BW Cir          & 2.54451   &  279.0 $\pm$ 4.7 & 1062 $\pm$ 91  &53.4 $\pm$ 8.2  & & & 3 \\
XTE J1550-564   & 1.5420333 &  363.1 $\pm$ 6.0 & 1506 $\pm$ 151 &28.4 $\pm$ 5.8  & 
11.5$\pm$3.9 & 75$\pm$4 & 4  \\
N. Oph 77       & 0.5228    &  441.0 $\pm$ 6.0 & 1885 $\pm$ 189 &68.5 $\pm$ 15.1 & 
6.2$\pm$1.2 & 70$\pm$10& 5 \\
N. Mus 91       & 0.4326058 &  406.0 $\pm$ 7.0 & 1735 $\pm$ 92  &48.7 $\pm$ 6.5  & 
7.0$\pm$0.6 & 54$\pm$2 & 6,7 \\
GS 2000+25      & 0.3440915 &  519.9 $\pm$ 5.1 & 2207 $\pm$ 341 &22.4 $\pm$ 7.2  & 
5.5$-$8.8 & 58$-$74 & 8,9 \\
A0620-00        & 0.32301405&  437.1 $\pm$ 2.0 & 1966 $\pm$ 110 &36.4 $\pm$ 10.8 & 
6.6$\pm$0.3 & 51$\pm$1 & 10,11 \\
XTE J1650-500   & 0.3205    &  435.0 $\pm$ 30.0& 1866 $\pm$ 348 &8.7  $\pm$ 2.3 & & & 12 \\
N Vel 93        & 0.285206  &  475.4 $\pm$ 5.9 & 2082 $\pm$ 208 &33.7 $\pm$ 7.4 & & & 13 \\
XTE J1859+226   & 0.274     &  541.0 $\pm$ 70.0& 2361 $\pm$ 109 &129.1$\pm$ 17.8 & & &14 \\
GRO J0422+320   & 0.2121600 &  380.6 $\pm$ 6.5 & 1716 $\pm$ 202 &165.3$\pm$ 18.2 & & &15 \\
XTE J1118+480   & 0.1699337 &  708.8 $\pm$ 1.4 & 2850 $\pm$ 187 &86.7 $\pm$ 18.9 &
6.9$-$8.2 & 68$-$79 &16,17 \\
\cutinhead{Neutron Stars}
Cen X-4         & 0.6290522 &  144.6 $\pm$ 0.3 &  678 $\pm$ 48  &36   $\pm$ 9  & 
1.94$^{+0.37}_{-0.85}$ & 32$^{+8}_{-2}$ & 18,19 \\
XTE J2123-058   & 0.24821   &  298.5 $\pm$ 6.9 & 1437 $\pm$272  &18.5 $\pm$5.0  & 
1.5$\pm$0.3 & 73$\pm$4 & 20,21 \\

\enddata
\tablenotetext{\dagger}{~This column provides references for the adopted values of 
$P$, $K_2$, $M_1$ and $i$.}
\tablerefs{
(1) \cite{casares96}; (2) \cite{khargharia10}; (3) \cite{casares09a}; (4) \cite{orosz11}; (5) \cite{harlaftis97}; 
(6) \cite{orosz96}; (7) \cite{gelino01};  (8) \cite{harlaftis96}; (9) \cite{ioannou04}; 
(10) \cite{gonzalez10}; (11) \cite{cantrell10}; 
(12) \cite{orosz04}; (13) \cite{filippenko99}; (14) \cite{corral11}; (15) \cite{webb00}; 
(16) \cite{gonzalez12}; (17) \cite{khargharia13};
(18) \cite{casares07}; (19) \cite{shahbaz14}; 
(20) \cite{tomsick01}; (21) \cite{zurita00}. }
\end{deluxetable}

\clearpage

\begin{deluxetable}{llrrrc}
\tablewidth{0pt}
\tablecaption{Cataclysmic Variables\label{tbl-4}}
\tablehead{
\colhead{Object} & \colhead{$P$ (d)} & \colhead{$K_{2}$ (km/s)} & 
\colhead{$FWHM$ (km/s)} & \colhead{$EW$ (\AA)} & \colhead{ref~\tablenotemark{\dagger}} }
\startdata
GK Per                & 1.9968     & 120.5$\pm$ 0.7   &  722 $\pm$  113 &  18.1$\pm$ 4.0 & 1 \\ 
SDSS 2044-04         & 1.68      &     90.0$\pm$ 8.0  &   420 $\pm$ 29 & 21.0 $\pm$  2.9 & 2 \\
BV Cen               & 0.611179  & 137.3$\pm$ 0.3   &  820  $\pm$ 58  &  13.0  $\pm$ 1.4  & 3-4 \\	
RX 1951.7+3716       & 0.492     & 81.0 $\pm$ 7.0   &   566 $\pm$ 40  &   29.0 $\pm$  4.1 & 2 \\
UY Pup               & 0.479269   & 102.0$\pm$ 4.0   &  615 $\pm$ 43 &  17.0 $\pm$ 2.4  & 5 \\
EY Cyg               & 0.4593249 &  54.0$\pm$ 2.0   &  341 $\pm$ 63 &   8.0 $\pm$ 3.0 & 5 \\
DX And               & 0.4405019   & 105.8$\pm$ 3.8   &  761 $\pm$ 96 &   5.4 $\pm$ 0.9 & 7-8 \\
SY Cnc               & 0.3823753  &  88.0$\pm$ 2.9   &  560 $\pm$ 58 &  10.9 $\pm$ 4.2 & 9  \\
AT Ara                 & 0.37551    &  99.5$\pm$ 3.2   &  618 $\pm$ 43 &  30.0 $\pm$ 4.2  & 10 \\
RU Peg               & 0.3746     & 121.0$\pm$ 2.0   &  881 $\pm$ 21 &  21.4 $\pm$ 01.4  & 11 \\
GY Hya              & 0.3472309   & 176.0$\pm$8.0    &  1267$\pm$  89 &  9.0  $\pm$ 1.3 & 2 \\
CH UMa              & 0.3431843  &  76.0$\pm$ 3.0   &  518 $\pm$ 22 &  46.3 $\pm$ 2.2  & 5 \\
V392 Hya            & 0.324952   &  144.0$\pm$ 9.0   &  850$\pm$  60 & 14.0 $\pm$  2.0 & 2 \\
RY Ser                & 0.3009     &  87.0$\pm$ 6.0   &  519 $\pm$ 36 &  22.0 $\pm$ 3.1  & 5 \\
HS 0218+3229   & 0.297229661  & 162.4$\pm$ 1.4   & 1023 $\pm$ 70 &  27.9 $\pm$ 4.0  & 12 \\
EM Cyg               & 0.290909    & 202.0$\pm$ 3.0   & 1242 $\pm$ 86 &  18.0 $\pm$ 0.9  &  13-14 \\
Z Cam                 & 0.289840    & 193.0$\pm$ 17.0  & 1107 $\pm$ 64 &  43.1 $\pm$ 1.3 & 15 \\
SDSS 0813+45   & 0.2890     &  54.0$\pm$ 7.0   &  461 $\pm$ 32 &  30.0 $\pm$ 4.2  & 5 \\
V426 Oph            & 0.285314  & 179.0$\pm$ 2.0   & 1224 $\pm$ 13 &  12.5 $\pm$ 0.3 & 16-17 \\
SS Cyg                & 0.27512973  & 165.0$\pm$ 1.0   &  986 $\pm$  14 &  39.5 $\pm$ 2.0  & 17-18 \\
LY UMa                 & 0.271278   & 141.0$\pm$ 3.0   & 1002 $\pm$ 70 & 28.0 $\pm$ 5.6 & 19 \\
BF Eri                  & 0.270881    & 182.5$\pm$ 0.9   & 1069 $\pm$ 34 &  27.7 $\pm$ 0.5 & 20 \\
TT Crt                  & 0.2683522   & 212.0$\pm$  5.0 & 1310 $\pm$ 92 &  17.0 $\pm$ 2.4  & 5 \\
AH Her                & 0.258116    & 175.0$\pm$ 2.0   & 1037 $\pm$ 52 &  16.2 $\pm$ 1.9  & 18,21 \\
EX Dra                & 0.20993698 & 210.0$\pm$ 14.0  & 1292 $\pm$ 63 &  37.0 $\pm$ 4.0 & 22-23 \\
HS 2325+8205    & 0.194334535 & 237.0$\pm$ 28.0  & 1377 $\pm$ 93 &  48.8 $\pm$ 7.2  & 24  \\
DQ Her                & 0.193620919 & 227.0$\pm$ 10.0  & 1168 $\pm$ 101 &  28.8 $\pm$ 1.5  & 25-26 \\
SDSS J100658+233724& 0.18591324 & 258.0$\pm$ 12.0  & 1446 $\pm$ 144 &  91.7 $\pm$ 10.2  & 27 \\
SS Aur                & 0.1828     & 167.0$\pm$ 15.0   &  911 $\pm$ 30 &  95.8 $\pm$ 5.2  & 28-29  \\
U Gem                 & 0.17690619  & 298.4 $\pm$ 9.0   & 1602 $\pm$ 53 & 87.7 $\pm$ 15.6  & 30-31 \\
CN Ori                & 0.16319006   & 231.0$\pm$ 7.0   & 1725 $\pm$395 &  23.8 $\pm$ 8.6 & 31-32 \\
IP Peg                & 0.1582061029  & 331.3$\pm$ 5.8   & 1899 $\pm$ 208 &  99.8 $\pm$ 28.0  & 33-34 \\
CTCV J1300-3052 & 0.088940717 & 378.0$\pm$ 6.0  & 2064 $\pm$ 66 & 127.2 $\pm$ 14.7  & 35-36 \\
QZ Ser                & 0.0831612 & 207.0$\pm$ 5.0 & 1218 $\pm$ 85 & 36.0 $\pm$ 5.0 & 37 \\
Z Cha                  & 0.0744992335  & 398.0$\pm$ 9.0 & 2000 $\pm$ 140 &  79.3  $\pm$ 11.1  & 38-39 \\
HT Cas                & 0.0736472029 & 389.0$\pm$ 9.0   & 1912 $\pm$ 155 & 154.4 $\pm$ 18.9  & 40-41 \\
OY Car                & 0.0631209343 & 470.0$\pm$ 2.7   & 2125 $\pm$ 147 & 73.7  $\pm$ 9.0 & 42-43 \\
V2051 Oph          & 0.0624278634 & 436.0$\pm$ 11.0  & 1981 $\pm$ 107 & 159.1 $\pm$ 14.2 & 44-45 \\
SDSS 103533.02+055158.3 & 0.0570067 & 520.0$\pm$3.0 & 1996 $\pm$ 125 & 51.5 $\pm$ 8.3 & 46  \\
WZ Sge              & 0.056688   & 510.0$\pm$ 15.0  & 1986 $\pm$ 129 & 150.1 $\pm$ 22.8 & 47-48 \\
SDSS J143317.78+101123.3 & 0.054240679 & 511.0$\pm$2.0 & 2460 $\pm$148 & 141.9$\pm$14.1 & 46 \\
SDSS 1507+52        & 0.04625834 & 529.0$\pm$17.0 & 2325 $\pm$163 & 87.0$\pm$17.4 & 49 \\
EI Psc = J2329+0628  & 0.044566 & 348.0$\pm$ 4.0 & 1865 $\pm$ 131 &  25.0 $\pm$ 3.5  & 50\\
\enddata
\tablenotetext{\dagger}{~This column provides references for the adopted values of $P$ and $K_2$.}
\tablerefs{(1) \cite{morales-rueda02}; 
(2) \cite{peters05}; 
(3) \cite{gilliland82}; (4) \cite{watson07}; (5) \cite{thorstensen04}:  
(6) \cite{echevarria07}; (7) \cite{bruch97}; (8) \cite{drew93} ; (9) \cite{casares09b}; (10) \cite{bruch03}; 
(11) \cite{stover81}; (12) \cite{rodriguez-gil09}; (13) \cite{robinson74}; (14) \cite{north00}; (15) \cite{kraft69}; 
(16) \cite{hessman88}; (17) \cite{north02}; (18) \cite{hessman84}; (19) \cite{tappert01};  
(20) \cite{neustroev08};  (21) \cite{horne86}; 
(22) \cite{baptista00}; (23) \cite{billington96}; (24) \cite{pyrzas12}; (25) \cite{wood05}; (26) \cite{horne93}; 
(27) \cite{southworth09}; (28) \cite{shafter86}; (29) \cite{davey92}; (30) \cite{marsh90}; (31) \cite{friend90}; 
(32) \cite{barrera89}; (33) \cite{copperwheat10}; (34) \cite{beekman00}; (35) \cite{savoury11}; (36) \cite{savoury12}; 
(37) \cite{thorstensen02a}; (38) \cite{wood86}; (39) \cite{wood90}; (40) \cite{borges08} ; (41) \cite{horne91}; 
(42) \cite{greenhill06}; (43) \cite{copperwheat12}; (44) \cite{baptista03}; (45) \cite{baptista98}; 
(46) \cite{littlefair08}; (47) \cite{patterson98}; (48) \cite{steeghs07}; (49) \cite{patterson08}; 
(50)\cite{thorstensen02b}.} 
\end{deluxetable}
\end{document}